\documentclass[reprint, aps, pra, superscriptaddress]{revtex4-2}

\usepackage{amsmath}
\usepackage{amssymb}
\usepackage{comment}
\usepackage{physics}
\usepackage[colorlinks=true]{hyperref}
\usepackage{graphicx}
\usepackage{soul}
\usepackage{bm}
\usepackage{graphicx,subcaption}
\usepackage{xcolor, mathtools}
\usepackage{color, soul}
\sethlcolor{red}

\begin{document}

\title{Radiation of breathing vortex electron packets in magnetic field}

\author{G.\,V.~Zmaga}
\email{george.zmaga@mail.ru}
\affiliation{School of Artificial Intelligence and Automation, Huazhong University of Science and Technology, Wuhan 430074, China}
\affiliation{School of Physics and Engineering,
ITMO University, 197101 St. Petersburg, Russia}

\author{G.\,K.~Sizykh}
\email{georgii.sizykh@metalab.ifmo.ru}
\affiliation{School of Physics and Engineering,
ITMO University, 197101 St. Petersburg, Russia}

\author{D.\,V.~Grosman}
\email{dmitriy.grosman@metalab.ifmo.ru}
\affiliation{School of Physics and Engineering,
ITMO University, 197101 St. Petersburg, Russia}

\author{Qi Meng}
\email{mengq8@mail2.sysu.edu.cn}
\affiliation{Sino-French Institute of Nuclear Engineering and Technology, Sun Yat-Sen University, Zhuhai 519082, China}

\author{Liping Zou}
\email{zoulp5@mail.sysu.edu.cn}
\affiliation{Sino-French Institute of Nuclear Engineering and Technology, Sun Yat-Sen University, Zhuhai 519082, China}

\author{Pengming Zhang}
\email{zhangpm5@mail.sysu.edu.cn}
\affiliation{School of Physics and Astronomy, Sun Yat-sen University, Zhuhai 519082, China}

\author{D.\,V.~Karlovets}
\email{dmitry.karlovets@metalab.ifmo.ru}
\affiliation{School of Physics and Engineering,
ITMO University, 197101 St. Petersburg, Russia}

\begin{abstract}
    When a vortex electron with an orbital angular momentum (OAM) enters a magnetic field, its quantum state is described with a nonstationary Laguerre-Gaussian (NSLG) state rather than with a stationary Landau state. A key feature of these NSLG states is oscillations of the electron wave packet's root-mean-square (r.m.s.) radius, similar to betatron oscillations. Classically, such an oscillating charge distribution is expected to emit photons. This raises a critical question: does this radiation carry away OAM, leading to a loss of the electron's vorticity? To investigate this, we solve Maxwell's equations using the charge and current densities derived from an electron in the NSLG state. We calculate the total radiated power and the angular momentum of the emitted field, quantifying the rate at which a vortex electron loses its energy and OAM while propagating in a longitudinal magnetic field. We find both the radiated power and the angular momentum losses to be negligible indicating that linear accelerators (linacs) appear to be a prominent tool for maintaining vorticity of relativistic vortex electrons and other charged particles, at least in the quasi-classical approximation.
\end{abstract}

\maketitle

\section{Introduction}

Vortex, or twisted, electrons are electrons in specific quantum states with a definite projection of an orbital angular momentum (OAM) along their propagation axis \cite{Bliokh2007, Bliokh2012, Bliokh2017}. Their OAM results in a magnetic moment that can significantly exceed its spin-originated counterpart. This property makes vortex electrons promising for numerous applications in material science, chemistry, and particle physics \cite{Grillo2017, Bliokh2017, Streshkova2024Nov}. However, in particle physics, specifically, electrons are required with energies in the GeV range, whereas the highest energy achieved for a vortex electron to date is only about 300 keV \cite{Uchida2010, Verbeeck2010, Bliokh2017}.

A straightforward approach to obtaining high-energy vortex electrons is to first generate twisted electrons at low energies and then accelerate them. While the curvilinear sections of accelerators are likely unsuitable since they lack axial symmetry, linear accelerators (linacs) appear to be appropriate for this purpose. However, a critical question arises: do these particles preserve their vorticity as they travel through the accelerator's electromagnetic fields? Specifically, could the small but inevitable radiation losses in a linac lead to a significant decrease in the particle's OAM?

Here we make a very first step towards addressing this question by employing a semiclassical approach for finding the radiated power and the corresponding loss of angular momentum. Our method begins with the wavefunction of a vortex electron, from which we derive the probability density and current. Multiplying by the electric charge and interpreting these densities as the effective distribution of the electric charge and current, we then use them as the source terms on the right-hand side of Maxwell's equations.

We model the linac as a region with homogeneous, longitudinal electric and magnetic fields. The primary role of the electric field is acceleration, and its contribution to radiation is negligible. Since a typical spatial coherence length of electron packets in accelerators does not exceed a few nanometers \cite{ehberger2015highly, cho2004quantitative, cho2013electron}, which is several orders of magnitude less than the spatial inhomogeneities of the accelerating electric field and focusing magnetic field, the packet locally probes the homogeneous fields. Therefore, our analysis of radiation focuses exclusively on the effects of the longitudinal magnetic field.

A vortex electron in free space is most realistically described with a free nonstationary Laguerre-Gaussian ($\text{NSLG}_{\text{f}}$) state \cite{McMorran2011, Schachinger2015, Bliokh2017, McMorran2017, Silenko2019, Sizykh2024Apr, Sizykh2024AprTransmission, Sheremet2025May}. Due to its nonstationary dynamics, the electron cannot be described with a stationary Landau state upon entering the magnetic field. Instead, its motion is accurately characterized with the nonstationary counterparts of Landau states: NSLG states in a magnetic field \cite{Zou2021Jan, Karlovets2021Vortex, Melkani2021, Zou2024Feb, Sizykh2024Apr, Sizykh2024AprTransmission}. Their distinguishing feature lies in oscillations of the state’s root-mean-square (r.m.s.) radius, resembling the phenomenon of betatron oscillations. Consequently, even within a magnetic field collinear with its average momentum, the electron’s wave packet behaves as a ``breathing'' charge cloud. From the perspective of classical electrodynamics, such an oscillating charge distribution must radiate, and calculating the characteristics of this radiation is the primary focus of this work. We illustrate our results in two experimentally relevant scenarios --- for electron microscopes and linacs. Examining both of these, it is shown that the losses of OAM are negligible in the vast majority of cases, thus enabling transport and focusing of vortex electrons, protons, or ions in the longitudinal magnetic fields.

Throughout this paper, we use CGS units and define the electron charge as $e < 0$.

\section{NSLG states}

In this section we briefly review NSLG states (a thorough analysis of them can be found in \cite{Sizykh2024Apr}). These are an orthogonal complete set of nonstationary \textit{exact} solutions to the Schr\"{o}dinger equation,
\begin{equation}
\label{eq:Schrodinger}
    i \hbar \partial_t\psi(\bm{r},t) = \frac{(\hat{\bm{p}} - e \bm{\mathcal{A}} / c)^2}{2m}\psi(\bm{r},t),
\end{equation}
where $\hat{\bm{p}} = -i \hbar \nabla$ and the external field vector potential $\bm{\mathcal{A}} = \bm{0}$ in free space and

\begin{equation}
    \bm{\mathcal{A}} = \displaystyle\frac{H r_\perp}{2}\hat{e}_{\phi}
    \label{eq:vp_ext_H}
\end{equation}
in a constant homogeneous longitudinal magnetic field.

In free space the transverse dynamics of a spatially localized vortex electron is described with an $\text{NSLG}_{\text{f}}$ state (see Sec. 2.3 in \cite{Sizykh2024Apr}),
\begin{equation}
\label{NSLG}
\begin{aligned}
    & \Psi_{n,l}(\bm{r_\perp},t) = N_\perp \frac{r_\perp^{|l|}}{\sigma^{|l|+1}(t)} L_n^{|l|} \left(\frac{r_\perp^2}{\sigma^2(t)}\right) \times  \\ 
    \exp & \left[il\phi - i\Phi_{\text{G}}(t) -  \frac{r_\perp^2}{2\sigma^2(t)}\left(1-i\frac{\sigma^2(t)}{\lambda_{\text{C}} \mathcal{R}(t)}\right)\right],
\end{aligned}
\end{equation}
where $\lambda_{\text{C}} = \hbar / (m c)$ is the reduced Compton wavelength, $n \in \{ 0, 1, 2 ... \}$ and $l \in \{ 0, \pm 1, \pm 2 ... \}$ are radial and orbital quantum numbers, $N_\perp = \sqrt{n!/\left(\pi \left(n+\abs{l}\right)!\right)}$ is the normalization constant, and the optical functions --- deviation $\sigma(t) = \sigma_{\text{f}}(t)$, curvature radius $\mathcal{R}(t) = \mathcal{R}_{\text{f}}(t) = \sigma\left(t\right)/\sigma'\left(t\right)$, and Gouy phase $\Phi_{\text{G}}(t) = \Phi_{\text{G}}^{\text{f}}(t) = \left(2n + \abs{l} + 1\right) \arctan{(t - t_g)/\tau_\text{d}}$ with $t_g$ being the generation time of a vortex electron and the diffraction time $\tau_\text{d}$.

An important feature of such a state is the r.m.s. radius,
\begin{equation}
\label{eq:RMSRad}
    \sqrt{\langle r_\perp^2 \rangle - \langle \bm{r_\perp} \rangle^2}_\text{f} = \varrho_{\text{f}}(t) = \sigma_{\text{f}}(t)\sqrt{2n + |l| + 1}.
\end{equation}
Hence, when such a vortex electron approaches a solenoid from free space it spreads, which prevents it from continuously transitioning into a stationary Landau state \cite{Sizykh2024AprTransmission}.

Nonetheless, a continuous state can be constructed by considering NSLG states within the field instead of Landau states. They are solutions to Eq. \eqref{eq:Schrodinger} with the vector potential in the symmetric gauge (see Eq. \eqref{eq:vp_ext_H}). NSLG states in the magnetic field have the same form \eqref{NSLG} as in free space albeit with different optical functions $\sigma(t)$, $\mathcal{R}(t)$, and $\Phi_{\text{G}}(t)$. The r.m.s. radius is still determined by general Eq. \eqref{eq:RMSRad}. Its evolution is governed by the deviation function $\sigma(t)$, which is now given by
\begin{equation}
\label{eq:OptSolS}
    \begin{aligned}
        & \sigma(t) = \sigma_{\text{st}} \sqrt{1 + \sqrt{1 - \left( \frac{\sigma_{\text{L}}}{\sigma_{\text{st}}} \right)^4} \sin{ \left( \mathfrak{s} \omega_{\text{c}}t - \vartheta \right) }}, \\
        & \sigma_{\text{st}}^2 = \frac{\sigma_0^2}{2} \left( 1 + \left( \frac{\sigma_{\text{L}}}{\sigma_0} \right)^4 + \left( \frac{\sigma'_0 \sigma_{\text{L}}^2}{c \lambda_C \sigma_0} \right)^2 \right), \\
        & \vartheta = \arcsin{\frac{1 - (\sigma_0 / \sigma_{\text{st}})^2}{\sqrt{1 - \left( \sigma_{\text{L}} / \sigma_{\text{st}} \right)^4}}}.
    \end{aligned}
\end{equation}
Here $\sigma_0$ and $\sigma'_0$ are initial values of the wave packet's width deviation and its derivative that are dictated by the state of the incident free-space wave packet; $\omega_{\text{c}} = e H / (m c)$ is the cyclotron frequency, and

\begin{equation}
    \sigma_{\text{L}} = \sqrt{\frac{2 \hbar c}{\abs{e H}}}
\end{equation}
is the r.m.s. radius of the Landau state with $n = l = 0$. The deviation function $\sigma(t)$ indicates the wave packet's transverse size oscillates with the cyclotron frequency. The sign function

\begin{equation}
    \mathfrak{s} \equiv \mathfrak{s}(\sigma_0, \sigma_0') =
    \begin{cases}
        \text{sign}(\sigma_0'), \sigma_0' \neq 0,\\
        \text{sign}(\sigma_{\text{L}} - \sigma_0), \sigma_0' = 0 \; \text{and} \; \sigma_0 \neq \sigma_{\text{L}},\\
        0, \sigma_0' = 0 \; \text{and} \; \sigma_0 = \sigma_{\text{L}};
    \end{cases}
    \label{eq:signum}
\end{equation}
determines whether the wave packet initially expands or contracts. The latter scenario is a very special case when the NSLG state becomes a stationary Landau state. The Gouy phase $\Phi_{\text{G}}(t)$ is given by Eq. (29) in \cite{Sizykh2024Apr}, while $\mathcal{R}(t)$ in Eq. \eqref{NSLG} can be derived from the first line of Eq. (10) in \cite{Sizykh2024Apr} using the dispersion equations \eqref{eq:OptSolS}.

The electron's OAM is independent of its longitudinal motion, which is unaffected by the vacuum-solenoid boundary crossing. Thus, we choose it to be a simple Gaussian wave packet with the average longitudinal momentum $p_0$ and the ``size'' $\sigma_z$ \eqref{appa:longitudinal_wf} short enough that the electron dynamics remain unchanged when passing the transition region, as detailed in the Supplemental Material of Ref. \cite{Sizykh2024AprTransmission} (reference [29] there). We present the longitudinal wave function in Eq. \eqref{appa:longitudinal_wf}. We also note that $p_0$ does not contribute to further results.

The wave function \eqref{NSLG} with the deviation function \eqref{eq:OptSolS} yields the charge and current densities,
\begin{equation}
\label{eq:sources}
    \begin{aligned}
        & \rho(\bm{r}, t) = e \abs{\Psi\left(\bm{r}, t\right)}^2, \\
        & \bm{j} \left(\bm{r}, t\right) = e \Re{ \Psi^* \left(\bm{r}, t\right) \frac{\hat{\bm{p}}}{m} \Psi\left(\bm{r}, t\right)} - \frac{e^2}{m c} \abs{\Psi\left(\bm{r}, t\right)}^2 \bm{\mathcal{A}},
    \end{aligned}
\end{equation}
the exact expressions for which are given in Eq. \eqref{appa:densities}.

\newpage

\section{Radiation of NSLG states in\\
a uniform external magnetic field}

\begin{figure*}
    \centering
    \includegraphics[width=0.85\linewidth]{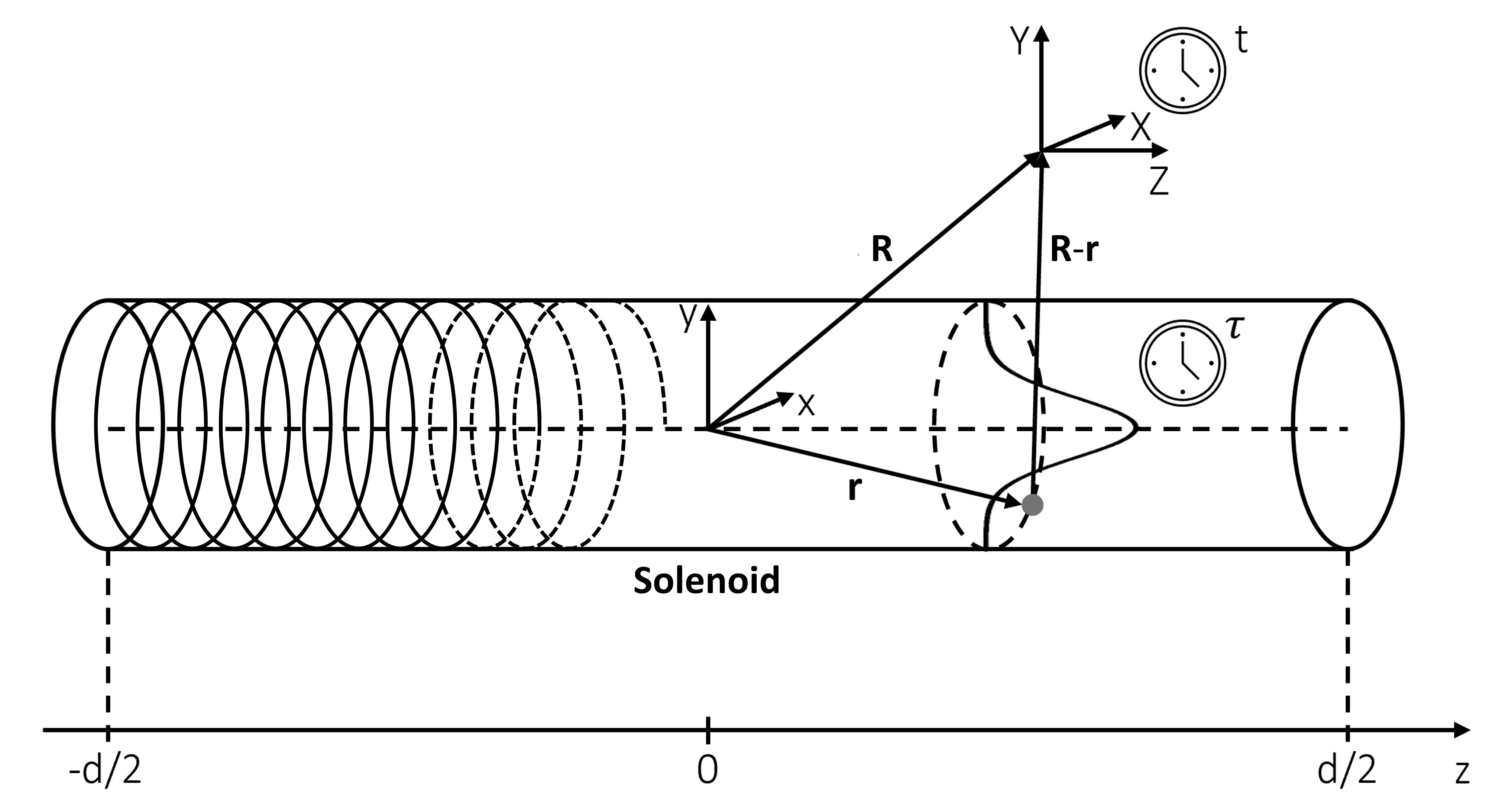}
    \caption{Schematic of the physical setup. A vortex electron wave packet enters a solenoid of length $d$. As it propagates along the $z$-axis, its wave function oscillates (the ``breathing'' NSLG mode). This non-stationary charge distribution emits radiation. The field emitted from a source point $\bm{r}$ at an earlier time $\tau$ is observed at a distant point $\bm{R}$.}
    \label{fig:ref_frame_scheme}
\end{figure*}

\subsection{Characteristics of emitted radiation}

From a classical electrodynamics perspective, Eqs. \eqref{eq:sources} describe a non-uniformly moving charge cloud, which must emit electromagnetic waves. The corresponding scalar $\varphi(\bm{R}, t)$ and vector $\bm{A}(\bm{R}, t)$ potentials of the field are
\begin{equation}
\label{PotsGen}
    \begin{aligned}
        \displaystyle & \varphi \left(\bm{R}, t \right) = \int\limits_{\mathbb{R}^3} \frac{\rho \left(\bm{r}, t'\right)}{|\bm{R} - \bm{r}|} \dd^3 r,\\
        \displaystyle & \bm{A} \left(\bm{R}, t \right) = \frac{1}{c}  \int\limits_{\mathbb{R}^3} \frac{\bm{j} \left(\bm{r}, t'\right)}{|\bm{R} - \bm{r}|} \dd^3 r,
    \end{aligned}
\end{equation}
where $|\bm{R} - \bm{r}|$ is the distance between a point of the charge cloud with the radius-vector $\bm{r}$ and the observation point at $\bm{R}$ and $t' = t - |\bm{R} - \bm{r}| / c$ is the retarded time. The setup it schematically depicted in Fig. \ref{fig:ref_frame_scheme}.

At this point, we employ the standard far-field approximation, assuming that the characteristic size of the radiating system is much smaller than the distance to the observation point. However, in addition to the time-derivative term, we also take into account the one arising from the denominator expansion. This term provides the leading-order contribution to the angular momentum of radiation, although \textit{it is irrelevant for the radiation power}.

Accordingly, we expand integrands in Eq. \eqref{PotsGen} into a series:
\begin{equation}
\label{Expansion}
    \frac{f\left(\bm{r}, t'\right)}{|\bm{R} - \bm{r}|} \approx \frac{f\left(\bm{r},\tau\right)}{R} + \frac{\bm{r} \cdot \hat{e}_{\bm{R}} \; \partial_\tau f\left(\bm{r},\tau\right)}{R c} + \frac{\bm{r} \cdot \hat{e}_{\bm{R}} \; f\left(\bm{r},\tau\right)}{R^2},
\end{equation}
where $f$ is a charge density or the component of a current density, $\tau = t - R / c$, $\hat{e}_{\bm{R}} = \bm{R}/R$. For convenience, we put the origin of the coordinate system in the middle of a solenoid (inside a transmission electron microscope or a linac). Hence, $\abs{\bm{r}} < d/2$, where $d$ is the solenoid length.

Using the expansion \eqref{Expansion} we have calculated the potentials (see Eq. \eqref{appb:potentials}). To proceed further, we evaluate the electric and magnetic fields. Exact expressions are presented in the appendix (see Eq. \eqref{appb:fields} for derived fields). The fields can be represented as a sum:
\begin{equation}
    \begin{aligned}
        & \bm{F} = \bm{F}_{\text{far}} + \bm{F}_{\text{near}}, \\
        & \bm{F}_{\text{far}} \propto R^{-1}, \hspace{20pt} \bm{F}_{\text{near}} \propto R^{-2},
    \end{aligned}
\end{equation}
where $\bm{F} = \{ \bm{E}, \bm{H} \}$. The term $\bm{F}_{\text{far}}$ represents the radiated part of the electromagnetic field, while $\bm{F}_{\text{near}}$ corresponds to the near field of the wave packet.

Then, we find the Poynting vector,
\begin{equation}
\label{eq:Poynting_general}
    \bm{S} = \frac{c}{4\pi} \bm{E} \times \bm{H},
\end{equation}
which we will further use to calculate both the radiated power and angular momentum of the radiation. The exact expression shown in Eq. \eqref{appb:poynting_vector} indicates that the Poynting vector can be represented as the following sum:
\begin{equation}
\label{PoyntingVectorGeneral}
    \begin{aligned}
        & \bm{S} = \bm{S}_{\text{far}} + \bm{S}_{\text{int}} + \bm{S}_{\text{near}}, \\
        & \bm{S}_{\text{far}} \propto R^{-2}, \hspace{20pt} \bm{S}_{\text{int}} \propto R^{-3}, \hspace{20pt} \bm{S}_{\text{near}} \propto R^{-4}.
    \end{aligned}
\end{equation}
As we will see, the first term in the sum describes the radiation field and provides the leading contribution to the radiation power, while the second term is essential when considering angular momentum of radiation. The near-field term, $\bm{S}_{\text{near}}$, does not contribute to the radiated energy or angular momentum.

\begin{figure*}
    \centering
    \includegraphics[width=\linewidth]{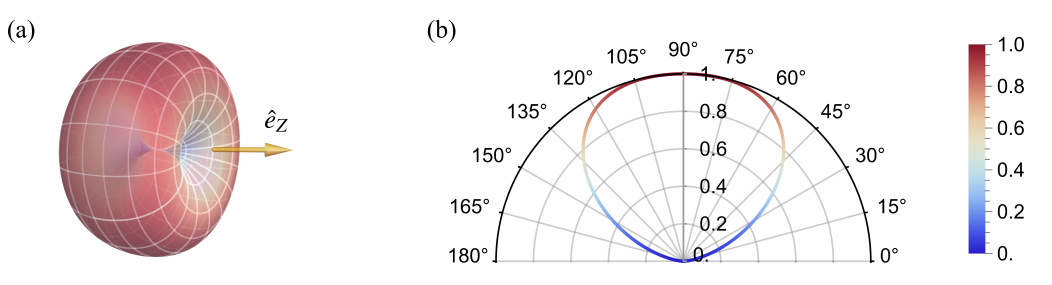}
    \caption{Angular distribution of the period-averaged radiation power $\langle \dd P / \dd \Omega \rangle_{T_{\text{c}}} \propto (1 + \cos^2{\theta}) \sin^2{\theta}$. (a) 3D spherical representation. (b) Polar plot with $\theta$ as the angular coordinate and the radial distance representing the distribution. Both plots share a common color scale indicating the distribution magnitude.}
    \label{fig:angular_ditribution}
\end{figure*}

\subsection{Power of radiation}

The power (energy per unit of time) radiated by an NSLG electron in a magnetic field is given by the integral
\begin{equation}
\label{eq:power_general}
    P\left( t \right) = \int\limits_{\mathbb{S}^2} \bm{S}\left( t \right) \cdot \dd \bm{s} \xrightarrow[R \rightarrow \infty]{\hspace{1cm}} \int\limits_{\mathbb{S}^2} \bm{S}_{\text{far}}\left( t \right) \cdot \dd \bm{s}
\end{equation}
over a large sphere $\mathbb{S}^2$ with the radius $R$, where $\dd \bm{s} = R^2 \dd \Omega \hat{e}_{\bm{R}}$ is the surface area of the infinitesimal sphere. The unit vector $\hat{e}_{R} = \bm{R} / R$ can be written using spherical coordinates: $\hat{e}_{R} = \sin{\theta}\hat{e}_{R_\perp} + \cos{\theta}\hat{e}_Z$. Unit vectors $\hat{e}_{R_\perp}$ and $\hat{e}_Z$ are directed along the transversal radius-vector $\bm{R_\perp}$ ($R_\perp = R \sin{\theta}$) and $z$-axis ($Z = R \cos{\theta}$), respectively. From Eq. \eqref{PoyntingVectorGeneral} we see that the only contribution to the power at distances $R \rightarrow \infty$ arises from $\bm{S}_{\text{far}}$, while the interference and near-field terms of the Poynting vector lead to vanishing contributions. Hence, the r.h.s. of Eq. \eqref{eq:power_general} holds.

The dynamics of an NSLG electron in a magnetic field are periodic in time with the cyclotron period

\begin{equation}
    T_{\text{c}} = \frac{2 \pi}{\abs{\omega_{\text{c}}}} = \frac{2 \pi m c}{\abs{e H}}.
\end{equation}
Therefore, we calculate the average power over one period:
\begin{equation}
\label{eq:power_averaged_general}
    \langle P \rangle_{T_{\text{c}}} = \frac{1}{T_{\text{c}}} \int\limits_{T_{\text{c}}} P\left( t \right) \dd t.
\end{equation}
The exact expression for $\langle P \rangle_{T_{\text{c}}}$ is given by Eq. \eqref{eq:angular_distribution_exact}. Figure \ref{fig:angular_ditribution} shows the angular distribution of the time-averaged radiated power, calculated from Eq. \eqref{eq:power_averaged_general}. The emission pattern is azimuthally symmetric and its dependence on the polar angle $\theta$ closely resembles that of classic dipole radiation.

\begin{figure*}
    \centering
    \includegraphics[width=1.\linewidth]{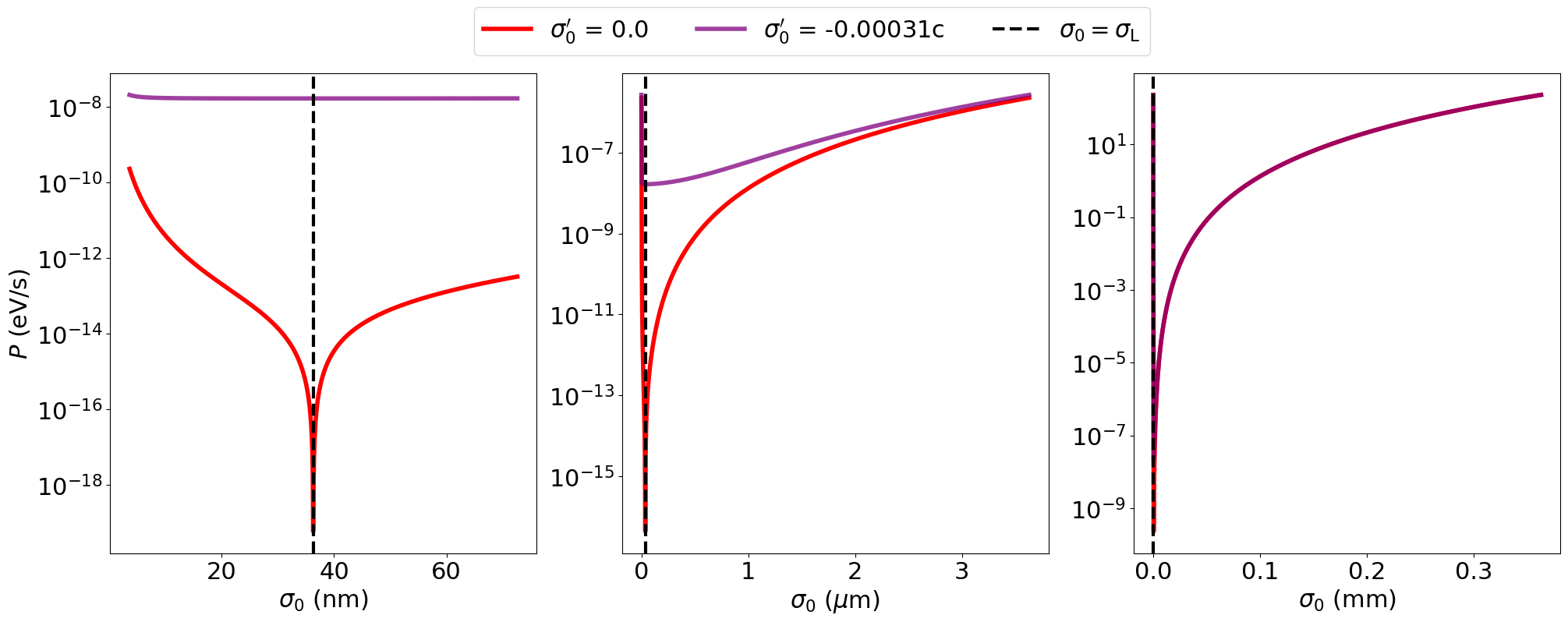}
    \caption{Period-averaged radiated power as a function of the initial wave packet's width deviation $\sigma_0$. The value of $\sigma_0' = -3.1 \times 10^{-4} c$ was taken from Ref. \cite{schattschneider2014imaging} that has been thoroughly considered in sec. V B of \cite{Sizykh2024Apr}. The following parameters are used: $H = 1$~T (corresponding to $\sigma_{\text{L}} \approx 36$ nm and $\hbar \omega_{\text{c}} \approx 10^{-4}$ eV), $n = 0$, $l = 10$. The panels show the dependence over three distinct scales of $\sigma_0$: (a) nanometers, (b) micrometers, and (c) sub-millimeters.}
    \label{fig:power_final}
\end{figure*}

Evaluating the integral in Eq. \eqref{eq:power_averaged_general} yields:
\begin{equation}
\label{Power}
    \displaystyle \langle P \rangle_{T_{\text{c}}} = (2n + \abs{l} + 1)^2 \frac{\mathfrak{s}^2\omega_{\text{c}}^6 e^2}{40 c^5} \left( \sigma_\text{st}^4 - \sigma_\text{L}^4 \right),
\end{equation}
where $\sigma_\text{st}$ --- the deviation of the period-averaged wave packet's width --- and $\sigma_L$ are taken from Eqs. \eqref{eq:OptSolS}. We demonstrate the results in Fig. \ref{fig:power_final} and discuss it in detail below in sec. \ref{subsec:analysis}.

\subsection{Angular momentum of radiation}

Next, we calculate the rate at which the electron's OAM is carried away by the radiation. The flux density of the angular momentum in the radial direction is given by
\begin{equation}
\label{AMFluxDensity}
    \bm{f} = \bm{l} c = \frac{1}{c} \bm{R} \cross \bm{S},
\end{equation}
where $\bm{l}$ is the angular momentum density. The rate of change of the field's angular momentum is then the integral of this flux density over a large sphere:
\begin{equation}
    \frac{\dd \bm{L}\left( t \right)}{\dd t} = \int\limits_{\mathbb{S}^2} \bm{f}\left( t \right) \dd s.
    \label{eq:OAM_decay}
\end{equation}
Decomposing the Poynting vector as in Eq. \eqref{PoyntingVectorGeneral}, we find that the rate of change of angular momentum also consists of three terms with different scaling properties:
\begin{equation}
    \begin{aligned}
        & \frac{\dd \bm{L}}{\dd t} = \frac{\dd \bm{L}}{\dd t}_{\text{far}} + \frac{\dd \bm{L}}{\dd t}_{\text{int}} + \frac{\dd \bm{L}}{\dd t}_{\text{near}}, \\
        & \frac{\dd \bm{L}}{\dd t}_{\text{far}} \propto R^1, \hspace{20pt} \frac{\dd \bm{L}}{\dd t}_{\text{int}} \propto R^0, \hspace{20pt} \frac{\dd \bm{L}}{\dd t}_{\text{near}} \propto R^{-1}.
    \end{aligned}
\end{equation}
At first glance, the far-field term appears to diverge as $R \rightarrow \infty$. However, this term is oscillatory and its time average over one cyclotron period \textit{vanishes} (see the text after Eq. \eqref{appb:oam_flux} for the proof).

Therefore, after time-averaging, the leading non-vanishing contribution to the net angular momentum loss arises solely from the \textit{interference term}:
\begin{equation}
    \Bigg\langle \frac{\dd \bm{L}}{\dd t} \Bigg\rangle_{T_{\text{c}}} = \frac{1}{T_{\text{c}}} \int\limits_{T_{\text{c}}} \frac{\dd \bm{L}\left( t \right)}{\dd t} _{\text{int}} \dd t.
\end{equation}
Thereby, we emphasize that the OAM losses are defined by the next order contribution, $1/R^3$, with respect to the energy emitted in the far field, $1/R^2$.

\begin{figure*}
    \centering
    \includegraphics[width=1.\linewidth]{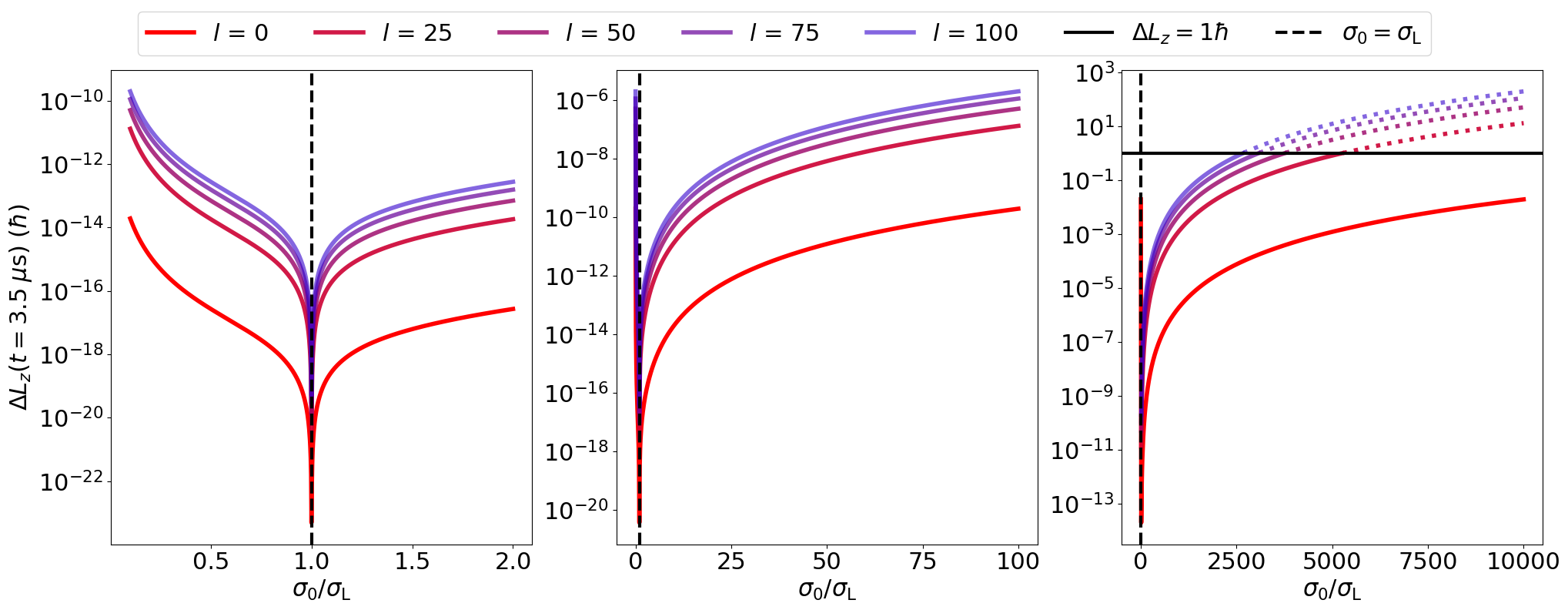}
    \caption{Period-averaged OAM loss per 1 km linac flight time $t = 3.5 \; \mu\text{s}$ as a function of the normalized initial wave packet's width deviation $\sigma_0/\sigma_\text{L}$. The following parameters are used: $H = 1$~T (corresponding to $\sigma_{\text{L}} \approx 36$ nm and $\hbar \omega_{\text{c}} \approx 10^{-4}$ eV), $\sigma'_0 = 0$, $n = 0$. The panels show the dependence over three distinct scales of $\sigma_0/\sigma_\text{L}$: (a) $\sigma_0 \approx \sigma_\text{L}$, (b) $\sigma_0 > \sigma_\text{L}$, and (c) $\sigma_0 \gg \sigma_\text{L}$. Horizontal line $\Delta L_z = 1 \hbar$ on the graph (c) shows the limitation of the utilized semiclassical approach that does not allow to predict angular momentum loss exceeding the single quantum of OAM.}
    \label{fig:oam_decay_final}
\end{figure*}

Evaluating this integral yields the time-averaged rate of the OAM lost by the electron:
\begin{equation}
\label{OAM}
    \displaystyle \Bigg\langle \frac{\dd \bm{L}}{\dd t} \Bigg\rangle_{T_{\text{c}}} = (2n + \abs{l} + 1)^2 \frac{\mathfrak{s}^2\omega_{\text{c}}^5 e^2}{120 c^5} \left( \sigma_\text{st}^4 - \sigma_\text{L}^4 \right) \hat{e}_Z.
\end{equation}
We demonstrate the results in Fig. \ref{fig:oam_decay_final} and elaborate on them in the following subsection.

\subsection{Analysis of results}
\label{subsec:analysis}

The final expressions for the radiated power, Eq. \eqref{Power}, and the rate of OAM loss, Eq. \eqref{OAM}, are structurally similar and provide valuable physical insights. First, we note that both expressions are proportional to $\mathfrak{s}^2$, where $\mathfrak{s}$ is defined in Eq. \eqref{eq:signum}. For stationary Landau states, $\mathfrak{s} = 0$, and consequently, there is no radiation. In our semiclassical model, such a stationary current configuration only generates a quasi-static near field.

For a general NSLG state ($\mathfrak{s} \neq 0$), the radiated power and rate of OAM loss are highly sensitive to the magnetic field, scaling as $\langle P \rangle_{T_{\text{c}}} \propto H^6$ and $\langle \dd \bm{L} / \dd t \rangle_{T_{\text{c}}} \propto H^5$. Note that $\langle \dd \bm{L} / \dd t \rangle_{T_{\text{c}}}$ is an odd function of the magnetic field's $z$-axis projection meaning the electron can both lose and gain OAM depending on the field direction. The dependence on the field is also notably steeper than the $P \propto H^4$ scaling of the  nonrelativistic synchrotron radiation \cite{STeng}. Importantly, in our setup the radiation characteristics can be tuned by varying the magnetic field strength independently, without the need to adjust the electron's energy to maintain its trajectory as is required in synchrotrons.

The power and OAM loss scale as $(2n + |l| + 1)^2$. Since vortex electrons typically have $|l| \gg n$, the radiation power and angular momentum are approximately proportional to $|l|^2$. Thus, particles with higher OAM radiate more intensely, although this usually implies a larger initial wave packet radius.

Finally, the radiation characteristics are proportional to the term $\sigma_{\text{st}}^4 - \sigma_{\text{L}}^4$, directly linking the emission to the initial conditions: wave packets that are larger or expand more rapidly upon entering the solenoid radiate more intensely.

To illustrate the magnitude of the radiated power and OAM loss rate, we use a typical set of parameters for a transmission electron microscope (TEM) in figures \ref{fig:power_final} and \ref{fig:oam_decay_final}: a longitudinal kinetic energy $E_{\parallel} = 200$~keV ($v \approx 0.7 c$), a $d = 20$ cm long solenoid, and an $H = 1$~T magnetic field ($\sigma_{\text{L}} \approx 36$ nm, $\hbar \omega_{\text{c}} \approx 10^{-4}$ eV). The value of the energy radiated over the flight time, evaluated from the power values in figure \ref{fig:power_final}, corresponds to a flight time $t_{\text{flight}} \equiv d / v \approx 1$~ns, or about 30 cyclotron periods ($T_{\text{c}} \approx 36$ ps). The quantum numbers are

\begin{equation}
    \begin{aligned}
        n &= 0,\\
        l &= 10,
    \end{aligned}
\end{equation}
if not stated otherwise. These parameters are also characteristic of a solenoid within a linac but with the electron's near-luminal velocity ($v \approx c$) being the primary difference. For a fixed solenoid length, this higher velocity reduces the total interaction time, resulting in slightly less total radiated energy and OAM.

As shown in Fig. \ref{fig:power_final}, the calculated radiated power for the TEM parameters is highly sensitive to the initial wave packet width. It scales from $10^{-8}$ eV/s for nanometer-scale widths up to $10^{1}$ eV/s for sub-millimeter widths. Despite this large range, the total radiated energy over the 20 cm path ($t_{\text{flight}} \approx 1$~ns) remains extremely small, reaching a maximum of only $10^{-5}$ eV. This value is more than an order of magnitude smaller than the cyclotron energy quantum ($\hbar \omega_{\text{c}} \approx 10^{-4}$ eV). This implies that for a typical TEM setup the probability of emitting even a single photon is very low, and the electron's energy loss is negligible. However, the situation changes dramatically for longer travel distances. For example, in a 1-km-long linac section ($t_{\text{flight}} \approx 3.5 \, \mu$s), an electron with similar transverse parameters would radiate a total energy of approximately $3.5 \cdot 10^{-2}$ eV. This energy corresponds to the emission of roughly 350 photons at the cyclotron frequency, indicating that the radiation becomes a significant physical effect.

Fig. \ref{fig:oam_decay_final} presents our central result: the calculated OAM loss. For the parameters considered, this rate varies dramatically with the initial wave packet size, from a minuscule $10^{-16} \hbar$ for nanometer-scale packets to a maximum of about $10^3 \hbar$ for sub-millimeter packets. To place this in a practical context, we can extrapolate these findings to a kilometer-scale linac. Even for a wide wave packet experiencing the maximum loss rate, the total OAM shed during a 1-km transit ($t_{\text{flight}} \approx 3.5 \, \mu$s) would be small for $\sigma_0 \gtrsim \sigma_\text{L}$: $\Delta L_z \approx 10^{-6} \hbar$. This loss is utterly negligible compared to the initial OAM of the vortex electron ($|l| \geq 1$). However, in case of extremely large $\sigma_0/\sigma_L$ OAM loss per the considered flight time becomes significant, reaching $1\hbar$ and exceeding this value. For the shown set of the orbital quantum numbers the smallest critical value of $\sigma_0$ where the semiclassical limit of the OAM radiation is overstepped is around $0.9$ mm ($l = 100$). It indicates that the semiclassical calculations do not describe the angular momentum loss process properly for large dispersion ratios, and the states of the wave packet changes. Based on this analysis, we come to a key conclusion: the loss of OAM for vortex electrons propagating through the magnetic fields typical of linacs is negligible for $\sigma_0/\sigma_\text{L} \gtrsim 1$. Therefore, \textit{linacs can be considered a robust and reliable platform for accelerating vortex electrons to relativistic energies}, at least within a quasi-classical approximation.

\begin{figure*}
    \centering
    \includegraphics[width=1.\linewidth]{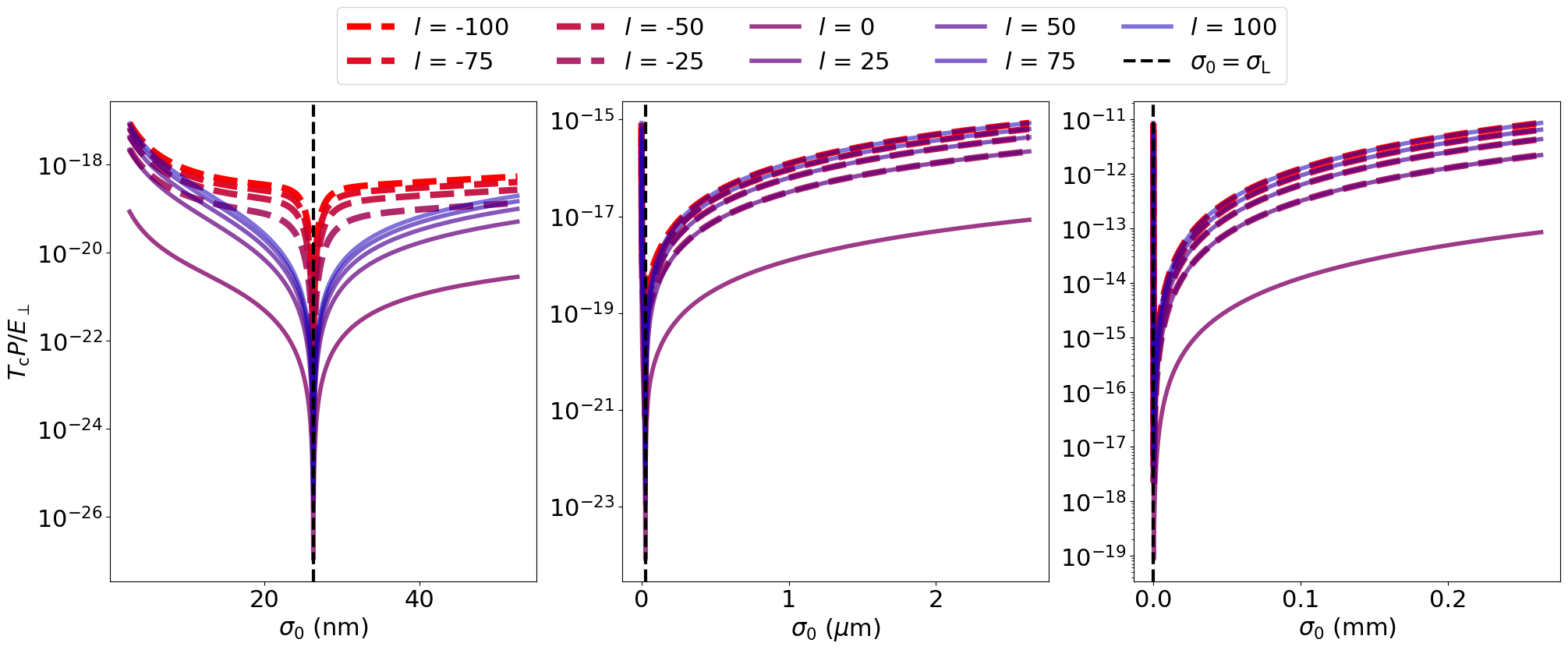}
    \caption{The ratio of the energy radiated for one cyclotron period to the energy of the electron's transverse motion, $E_{\text{rad}} / E_{\perp}$, as a function of the initial transversal size deviation $\sigma_0$ of the wave packet. The following parameters are used: $H = 1$~T (corresponding to $\sigma_{\text{L}} \approx 36$ nm and $\hbar \omega_{\text{c}} \approx 10^{-4}$ eV), $\sigma'_0 = 0$, $n = 0$. The panels show the dependence over three distinct scales of $\sigma_0$: (a) nanometers, (b) micrometers, and (c) sub-millimeters.}
    \label{fig:ratio_final}
\end{figure*}

\subsection{Applicability of semiclassical approach}

A key assumption of our semiclassical model is that the emission of very few soft photons does not change the quantum state of the electron much. Its wave function is assumed to evolve according to the Schrödinger equation with a fixed external field, and the back-action of the emitted radiation on this state is neglected.

To validate this assumption, we compare the energy radiated during one cyclotron period,
\begin{equation}
    E_{\text{rad}} = \langle P \rangle_{T_{\text{c}}} T_{\text{c}},
\end{equation}
with the expectation value of the electron's transverse energy,
\begin{equation}
    E_{\perp} = \frac{\hbar \omega_{\text{c}}}{2} (2n + \abs{l} + 1) \frac{\sigma_{\text{st}}^2}{\sigma_{\text{L}}^2} + l \mu_{\text{B}} H,
\end{equation}
where $\mu_{\text{B}} = \abs{e}\hbar/2 mc$ is the Bohr magneton. 

The semiclassical approach remains self-consistent as long as the radiated energy is a small fraction of the particle's transverse energy, i.e., $E_{\text{rad}} \ll E_{\perp}$. To verify this, we plot the ratio of the energy radiated per cyclotron period to the transverse energy in Fig. \ref{fig:ratio_final}. For the parameter set considered, this ratio is indeed much less than unity. We can extend this validity check to our most extreme scenario: a sub-millimeter wave packet traversing a 1-kilometer linac. Over the entire flight time ($t_{\text{flight}} \approx 3.5 \, \mu$s, or $\sim 10^5 T_{\text{c}}$), the total radiated energy still constitutes no more than $10^{-6}$ of the electron's transverse energy. This confirms that the back-action of radiation is negligible and robustly validates our semiclassical treatment even for kilometer-scale acceleration.

\section{Discussion}

Our semiclassical approach, which treats the electron's wave function as a coherent ``charge cloud'', is central to an ongoing debate in the literature on how to model single-particle radiation \cite{Remez2019Aug, Karlovets2021Jan, Karnieli2022Mar, Karlovets2022Mar}. An alternative perspective suggests an incoherent model, where radiation intensity is integrated over the probability distribution. Experimentally distinguishing between these competing theoretical frameworks — for instance, using a TEM — remains a crucial open challenge and a promising direction for future research.

Our semiclassical results must also be considered in the context of a full QED treatment, which includes spontaneous quantum transitions between energy levels -- a channel absent in our model. A more elaborate quantum analysis has recently been done by Murtazin et al. \cite{murtazin2025photon}, where a qualitative confirmation of our conclusions is presented. We hypothesize that a rigorous QED calculation for an NSLG state would reveal both this ``quantum'' radiation and the classical component from the wave packet's dynamics. This raises fundamental questions about how these two contributions are encoded within QED, whether they can be separated, and how they might interfere.

Plausibly, for ``sufficiently nonstationary'' states, the ``classical'' radiation mechanism we describe might dominate over the ``quantum'' one. A critical open question is the quantitative criterion for this classical regime, which is supposed to be a condition like $\sigma_{\text{st}} - \sigma_{\text{L}} \gg \sigma_{\text{L}}$.

A final practical consideration is the radiation from the solenoid's fringe fields. As detailed in Appendix \ref{AppC}, the inhomogeneous fields in this region generate a transient burst of radiation with possibly high values of angular momentum as the electron enters and exits. Crucially, this transition signal is experimentally distinguishable from the continuous, periodic radiation generated within the uniform field. This separation allows for a clear experimental test of our central predictions for the steady-state emission.

\section{Conclusion}

In this work, we have developed a semiclassical model to investigate the radiation emitted by a vortex electron propagating through a longitudinal magnetic field, such as in a linear accelerator. Our key finding is that both the radiated energy and the change in the electron's OAM are extremely small for realistic accelerator parameters in a very wide range of initial dispersion values of the wave packet. Even when extrapolated to a kilometer-scale linac, the total energy radiated remains a negligible fraction of the electron's transverse energy, which in turn validates our semiclassical approach. More importantly, the total change in OAM is found to be minuscule up to $\sigma_0 \gg \sigma_\text{L}$, far less than a single quantum of $\hbar$. In other words, we have found that a single quantum of OAM can be emitted during the flight time in a 1 km long linac, but for $\sigma_0 \gtrsim \sigma_\text{L}$ relatively small loss is present.

We therefore arrive at a clear and impactful conclusion: the orbital angular momentum of vortex electrons is remarkably robust during acceleration in linacs. This result strongly supports the viability of using linacs to produce high-energy, relativistic vortex electron beams for future applications in materials science and particle physics.

\section{Acknowledgment}

We are grateful to Alexandr Shchepkin, Igor Ivanov, and Ilia Pavlov for fruitful discussions and valuable comments on the applicability of the considered model to realistic scenarios. The study is supported by
the Russian Science Foundation (Project No. 23-62-10026; \href{https://rscf.ru/en/project/23-62-10026/}{https://rscf.ru/en/project/23-
62-10026/}). The work in Sec.\ref{subsec:analysis} was supported by the Foundation for the Advancement of Theoretical Physics and Mathematics “BASIS”.

\appendix

\section{Details on NSLG states}
\label{AppA}

The longitudinal wavefunction of the NSLG state is given by

\begin{equation}
\label{appa:longitudinal_wf}
    \begin{aligned}
        & \psi_{\parallel} = N_\parallel\int\limits_{-\infty}^{\infty} \dd p_z \exp \left\{i \left( \frac{p_z z}{\hbar} - \frac{p_z^2}{2m\hbar} t \right) - \frac{\left(p_z - p_0\right)^2}{2 \sigma_p^2}\right\} =\\
        & N_\parallel \frac{\sqrt{2 \pi} \hbar}{\sqrt[4]{\sigma_z^4 + \hbar^2 t^2 / m^2}} \exp\left\{-\frac{i}{2} \arctan{\frac{t}{t_{\text{d}}}} - \right.\\
        & \left. \frac{\tilde{z}^2(t)}{2 \sigma_z^2 \left(1 + i t / t_{\text{d}} \right)} + \frac{i p_0}{2\hbar}   \left(2\tilde{z}(t) + \frac{p_0 t}{m}\right)\right\}
    \end{aligned}
\end{equation}
with the classical trajectory $\tilde{z}(t) = z - p_0 t/m$, average momentum $p_0$, characteristic width $\sigma_z = \hbar / \sigma_p$, diffraction time $t_{\text{d}} = \sigma_z^2 m / \hbar = \sigma_z^2 / (c \lambda_{\text{C}})$, and the normalization coefficient $N_\parallel = \sigma_z^{1/2}/\left(\sqrt{2}\pi^{3/4} \hbar\right)$. This is just a Gaussian wave packet in momentum representation resulting in a Gaussian wave packet in coordinate representation with the waist dispersion $\sigma_z$ and spreading with time in the longitudinal direction.

The charge and current densities of the NSLG state determined by Eqs. \eqref{NSLG}, \eqref{eq:OptSolS}, and \eqref{appa:longitudinal_wf} are given by the following expressions:
\begin{widetext}
    \begin{equation}
            \begin{aligned}
                & \rho \left(\bm{r}, t\right) = \dfrac{2\pi \hbar^2 N_\parallel^2 N_\perp^2 e}{\sigma_z\sigma^2\left(t\right)\sqrt{\sigma_z^2 + \hbar^2 t^2/m^2\sigma_z^2}}  \left(\frac{r_\perp^2}{\sigma^2\left(t\right)}\right)^\abs{l}  \left(L_n^\abs{l}\left(\frac{r_\perp^2}{\sigma^2\left(t\right)}\right)\right)^2  \exp{-\frac{r_\perp^2}{\sigma^2\left(t\right)} -\dfrac{\tilde{z}^2(t)}{\sigma_z^2 + \hbar^2 t^2/m^2\sigma_z^2}},\\
                & \bm{j}\left(\bm{r}, t\right) = \frac{2\pi \hbar^3 N_\parallel^2 N_\perp^2 e r_\perp^{2\abs{l}-1}}{m \sigma_z \sqrt{\sigma_z^2 + \hbar^2 t^2/m^2\sigma_z^2}  \sigma^{2\abs{l}+2}\left(t\right)}  \left(L_n^\abs{l}\left(\frac{r_\perp^2}{\sigma^2\left(t\right)}\right)\right)^2 \left[\frac{r_\perp^2}{\lambda_C R\left(t\right)} \hat{e}_{R_\perp} + \left(l - \frac{ e H r_\perp^2}{2c \hbar}\right)\hat{e}_{\phi} \right. +\\
                & \left. r_\perp \left(\frac{\tilde{z}(t) t \hbar}{m\sigma_z^2\left(\sigma_z^2 + \hbar^2 t^2/m^2\sigma_z^2\right)} + \frac{p_0}{\hbar}\right)\hat{e}_Z\right]  \exp{-\frac{r_\perp^2}{\sigma^2\left(t\right)} - \frac{\tilde{z}^2(t)}{\sigma_z^2 + \hbar^2 t^2/m^2\sigma_z^2}},
            \end{aligned}
        \label{appa:densities}
    \end{equation}
\end{widetext}
where $N_{\perp} = \sqrt{n! / (\pi (n + |l|)! )}$ is the transverse normalization constant, the deviation $\sigma(t)$ is given by Eq. \eqref{eq:OptSolS}, and $\hat{e}_{R_\perp, \phi, Z}$  are the unit vectors of the cylindrical coordinate system.

\section{Field characteristics}
\label{AppB}

Within the expansion \eqref{Expansion}, the scalar and vector potentials determined by Eqs. \eqref{PotsGen} with the sources \eqref{appa:densities} defined by the state \eqref{NSLG} are:
\begin{widetext}
\begin{equation}
    \begin{aligned}
        \displaystyle & \varphi \left(\bm{R}, t\right) \approx \frac{e}{R} \left(1 + \frac{Z p_0 t}{R^2 m}\right),\\
        & \bm{A} \left(\bm{R}, t\right) \approx \frac{e\hbar}{2 R^2 m c^2} \left[\frac{R_\perp}{2\lambda_C c}  \left(\frac{c}{R}  \partial_t \varrho^2(t) + \partial^2_t \varrho^2(t)\right)\hat{e}_{R_\perp} + R_\perp \left\{\frac{l c}{R} - \frac{e H}{2\hbar c}  \left(\frac{c \varrho^2(t)}{R} + \partial_t \varrho^2(t)\right)\right\}\hat{e}_{\phi} \right. +\\
        & \left. \frac{2}{\hbar}\left\{\frac{Zct}{R m} \left(\frac{\hbar^2}{2\sigma_z^2} + p_0^2\right) + R p_0 c\right\}\hat{e}_Z\right],
    \end{aligned}
    \label{appb:potentials}
\end{equation}
\end{widetext}
where $\varrho^2(t) = (2n + |l| + 1) \sigma^2(t)$ is the mean square transverse radius of the NSLG wave packet.

The electric and magnetic components of the radiated field according to \eqref{appb:potentials} are, respectively,
\begin{widetext}
\begin{equation}
    \begin{aligned}
        & \bm{E} = \frac{e R_\perp}{R^2}  \left[\frac{1}{R} \left(1 + \frac{3Z p_0}{R mc}\right) - \frac{\hbar}{4\lambda_C m c^4}  \left(\frac{c}{R}  \partial^2_{t} \varrho^2(t) + \partial^3_{t} \varrho^2(t)\right)\right]\hat{e}_{R_\perp} + \frac{e^2 H R_\perp}{4 R^2 m c^4} \left(\frac{c}{R}  \partial_{t} \varrho^2(t) + \partial^2_{t} \varrho^2(t)\right)\hat{e}_{\phi} + \\
        & \frac{e}{R^3} \left[Z \left\{1 - \frac{1}{m^2 c^2} \left(\frac{\hbar^2}{2\sigma_z^2} + p_0^2\right)\right\} + \frac{\left(2Z^2 - R_\perp^2\right) p_0}{R mc}\right]\hat{e}_Z, \\
        & \bm{H} = -\frac{e^2 H R_\perp Z}{4 R^3 m c^4}  \left(\frac{3c}{R}  \partial_{t} \varrho^2(t) + \partial^2_{t} \varrho^2(t)\right)\hat{e}_{R_\perp} + \frac{e\hbar R_\perp}{2R^3 m c^4}  \left[-\frac{Z}{2\lambda_C}  \left(\frac{3c}{R}  \partial^2_{t} \varrho^2(t) + \partial^3_{t} \varrho^2(t)\right) + \frac{3\hbar Z c^2}{R m\sigma_z^2} + \right.\\
        & \left. \frac{2p_0 c^3}{\hbar} \left(1 + \frac{2Z p_0}{R mc}\right)\right]\hat{e}_{\phi} + \frac{e^2 H}{4 R^2 m c^3}  \left(\frac{3R_\perp^2}{R^2}  \partial_{t} \varrho^2(t) + \frac{R_\perp^2}{R c}  \partial^2_{t} \varrho^2(t) - 2 \partial_{t} \varrho^2(t)\right)\hat{e}_Z.
    \end{aligned}
    \label{appb:fields}
\end{equation}
\end{widetext}

According to Eqs. \eqref{eq:Poynting_general} and \eqref{PoyntingVectorGeneral}, the Poynting vector components are the following:
\begin{widetext}
\begin{equation}
\label{appb:poynting_vector}
        \begin{aligned}
            & \bm{S}_\text{far} = \frac{e^2 R_\perp^2}{64\pi R^5 c^5} \left[\omega_{\text{c}}^2 R_\perp \left(\partial^2_{t} \varrho^2(t)\right)^2 \hat{e}_{R_\perp} + \omega_{\text{c}} R_\perp \partial^3_{t} \varrho^2(t) \partial^2_{t} \varrho^2(t) \hat{e}_{\phi} + Z \left( \left(\partial^3_{t} \varrho^2(t)\right)^2 + \omega_{\text{c}}^2 \left(\partial^2_{t} \varrho^2(t)\right)^2 \right) \hat{e}_Z \right],\\
            & \bm{S}_\text{int} = \frac{e^2 R_\perp}{16\pi R^4 c^2} \left[ \left \{\frac{\omega_{\text{c}}^2 \partial^2_{t} \varrho^2(t) \partial_{t} \varrho^2(t)}{2 c^2}  \left(\frac{2 R_\perp^2}{R^2} - 1\right) + \frac{Z \partial^3_{t} \varrho^2(t)}{R^2}  \left(Z - \frac{\hbar^2 Z}{2\sigma_z^2 m^2 c^2} - \frac{\hbar p_0}{\lambda_C m^2 c^2} \left(\frac{R_\perp^2 - 2Z^2}{R} + \frac{Z p_0}{mc}\right)\right)\right\}\hat{e}_{R_\perp} + \right.\\
            & \left. \omega_{\text{c}} \left\{\frac{\partial^3_{t} \varrho^2(t) \partial_{t} \varrho^2(t)}{4 c^2}  \left(\frac{3 R_\perp^2}{R^2} - 2\right) - \partial^2_{t} \varrho^2(t) \left(1 - \frac{\hbar^2 Z^2}{2\sigma_z^2 R^2 m^2 c^2} + \frac{e H Z p_0}{R m^2 c^2 \omega_{\text{c}}} \left(2 - \frac{Z p_0}{R mc}\right) - \frac{R_\perp^2 \partial^2_{t} \varrho^2(t)}{4 R^2 c^2}\right)\right\}\hat{e}_{\phi} + \right.\\
            & \left.  \frac{R_\perp}{R} \left\{\partial^3_{t} \varrho^2(t) \left(\frac{Z\partial^2_{t} \varrho^2(t)}{R c^2} + \frac{3Z \sigma_p^2}{2 R m^2 c^2} - \frac{Z}{R} - \frac{\hbar p_0}{\lambda_C m^2 c^2} \left(1 + \frac{2Z p_0}{R mc} + \frac{3Z^2}{R^2}\right)\right) + \frac{\omega_{\text{c}}^2 Z \partial^2_{t} \varrho^2(t) \partial_{t} \varrho^2(t)}{R c^2}\right\}\hat{e}_Z \right].
        \end{aligned}
\end{equation}
\end{widetext}

The period-averaged angular distribution of the power determined by Eqs. \eqref{eq:power_general} and \eqref{eq:power_averaged_general} is given by
\begin{equation}
\label{eq:angular_distribution_exact}
\begin{aligned}
    & \Big\langle \frac{\dd P}{\dd \Omega} \Big\rangle_{T_{\text{c}}} = \\
    & (2n + \abs{l} + 1)^2 \frac{\mathfrak{s}^2\omega_{\text{c}}^6 e^2}{40 c^5} \left( \sigma_\text{st}^4 - \sigma_\text{L}^4 \right) (1 + \cos^2{\theta}) \sin^2{\theta}.
\end{aligned}
\end{equation}

The angular momentum flux density \eqref{AMFluxDensity} in the far field derived from the corresponding Poynting vector component takes the following form:
\begin{equation}
\label{appb:oam_flux}
    \begin{aligned}
        & \bm{f}_\text{far} = \frac{e^2 R_\perp^2}{64\pi R^5 c^6} \left[ - \omega_{\text{c}} R_\perp Z \partial^3_{t} \varrho^2(t) \partial^2_{t} \varrho^2(t) \hat{e}_{R_\perp} - \right. \\
        & \left. R_\perp Z \left(\partial^3_{t} \varrho^2(t)\right)^2 \hat{e}_{\phi} + \omega_{\text{c}} R_\perp^2 \partial^3_{t} \varrho^2(t) \partial^2_{t} \varrho^2(t) \hat{e}_Z\right].
    \end{aligned}
\end{equation}

To find the angular momentum rate of change we integrate Eq. \eqref{appb:oam_flux} over a large sphere. Its components directed along  $\hat{e}_{R_\perp, \phi}$ contain oscillating terms proportional to $\sin{\phi}, \; \cos{\phi}$ and vanish after integration over the solid angle. Only $z$-component remains and is determined by the following expression:
\begin{equation}
    \frac{\dd L_z}{\dd t}_\text{far} = \frac{e^2 R \omega_{\text{c}}}{30 c^6} \partial^3_t \varrho^2(t) \partial^2_t \varrho^2(t).
\end{equation}

According to the first line in Eq. \eqref{eq:OptSolS}, time derivatives of $\varrho^2(t)$ of odd orders are proportional to $\sin{(\mathfrak{s}\omega_{\text{c}} t - \vartheta)}$ and of even orders --- to $\cos{(\mathfrak{s}\omega_{\text{c}} t - \vartheta)}$. Thus, averaging terms $\partial^n_t \varrho^2(t) \partial^m_t \varrho^2(t)$ with $n$ and $m$ of different parity  over the cyclotron period results in zero. Hence, far fields do not contribute to the angular momentum rate of change.

\section{Influence of solenoid fringe fields on the radiation}
\label{AppC}

We consider a setup with a solenoid of diameter $D \sim 1$ cm and an internal field of $H = 1$ T. The corresponding cyclotron frequency and period are $\omega_{\text{c}} \approx 3.34 \times 10^{11}$ rad/s and $T_{\text{c}} \approx 18.8$ ps, respectively.

A relativistic electron traverses the fringe field region of length $2 D$ in a time $t_\text{trans} \sim 2D / c \approx 67$ ps. Since the transit time is longer than the cyclotron period ($t_\text{trans} \approx T_\text{c}$), the magnetic field experienced by the electron changes slowly over a single cyclotron period. This justifies the use of the adiabatic approximation, where the rate of change of the field is small compared to its characteristic frequency scales:
\begin{equation}
\left|\frac{\partial \omega_{\text{c}}}{\partial t}\right| \ll \frac{\omega_{\text{c}}}{T_c} \sim \omega_c^2.
\end{equation}

Under this condition, we can show that an approximate solution to the Schrödinger equation \eqref{eq:Schrodinger} remains an NSLG state \eqref{NSLG}, but with a time-dependent cyclotron frequency. This follows from the analysis of the equations for the optical functions, which retain their form with the magnetic length now being a function of time: $\sigma_\text{L} \rightarrow \sigma_\text{L}\left(t\right) = \sqrt{2\hbar c/\abs{e H\left(t\right)}}$.

To simplify the calculation, we model the change in the cyclotron frequency across the transition region as linear, i.e., $\omega_{\text{c}} (t) \approx \omega_{\text{c}}' t$, where the constant rate of change is $\omega_{\text{c}}' \approx \omega_{\text{c}} / t_{\text{trans}}$. Then, we repeat the derivation from \cite{Sizykh2024Apr}, substituting this time-dependent frequency into the system of optical equations [23] in that paper. The resulting expressions for the power and angular momentum components, averaged over the transit time $T \equiv t_{\text{trans}}$, are:

\begin{widetext}
\begin{equation}
    \begin{cases}
        \begin{split}
            &\displaystyle\langle P\rangle_{T} = (2n + \abs{l} + 1)^2 \frac{\omega_{\text{c}}^2 e^2}{12 T^2 c^5} (\sigma^4_\text{st} - \sigma^4_\text{L}) \left[\frac{4}{5} \left(\omega_{\text{c}} T\right)^2 + 1 + \frac{9}{16} \sqrt{\frac{\pi}{\omega_{\text{c}} T}} C\left(2\sqrt{\frac{\omega_{\text{c}} T}{\pi}}\right) + \frac{\mathfrak{s}-3}{8} \cos{(2 \omega_{\text{c}} T)} - \right.\\
            & \left. \frac{\omega_{\text{c}} T}{2} \sin{(2 \omega_{\text{c}} T)} \right],\\[5pt]
        \end{split}\\
        \begin{split}
            & \displaystyle \bigg{\langle}\frac{\dd \bm{L}}{\dd t}\bigg{\rangle}_{\text{int}, T} = \frac{\omega_{\text{c}} e^2}{4 c^3} \left[ (2n + \abs{l} + 1) \frac{2\mathfrak{s} \omega_{\text{c}}}{3 T} \sqrt{\sigma^4_\text{st} - \sigma^4_\text{L}} \left(\frac{\hbar^2}{5\sigma_z^2 m^2 c^2} - 2\right) \cos{(\omega_{\text{c}} T)} +  \right.\\
            & \left. (2n + \abs{l} + 1)^2 \frac{\omega_{\text{c}}^2}{15 T^2 c^2} (\sigma^4_\text{st} - \sigma^4_\text{L}) \left\{\frac{4}{5} \left(\omega_{\text{c}} T\right)^2 + 1 + \frac{9}{16} \sqrt{\frac{\pi}{\omega_{\text{c}} T}} C\left(2\sqrt{\frac{\omega_{\text{c}} T}{\pi}}\right) + \frac{\mathfrak{s}-3}{8} \cos{(2\omega_{\text{c}} T)} - \right. \right.\\
            & \left. \left. \frac{\omega_{\text{c}} T}{2} \sin{(2\omega_{\text{c}} T)} \right\} \right] \hat{e}_Z,\\[5pt]
        \end{split}\\
        \begin{split}
            \displaystyle \bigg{\langle}\frac{\dd \bm{L}_{p_0}}{\dd t}\bigg{\rangle}_{\text{int}, T} &= (2n + \abs{l} + 1) \frac{\mathfrak{s}\pi\omega_{\text{c}}^2 e^2 p_0^2}{64 T m^2 c^4} \sqrt{\sigma^4_\text{st} - \sigma^4_\text{L}} \cos{(\omega_{\text{c}} T)} \hat{e}_Z,\\[5pt]
        \end{split}\\
        \begin{split}
            \displaystyle \bigg{\langle}\frac{\dd \bm{L}}{\dd t}\bigg{\rangle}_{\text{rad}, T} &= (2n + \abs{l} + 1)^2 \frac{5\pi \omega_{\text{c}}^3 e^2 R}{256 T^3 c^6} (\sigma^4_\text{st} - \sigma^4_\text{L}) \left[\omega_{\text{c}} T \sin{(2\omega_{\text{c}} T)} + \left(1 - 2\omega_{\text{c}}^2 T^2\right)\sin^2{(\omega_{\text{c}} T)}\right] \hat{e}_Z.
        \end{split}
    \end{cases}
\end{equation}
\end{widetext}
where $C\left(x\right) = \displaystyle\int\limits_0^x \cos{\frac{\pi t^2}{2}} \dd t$ is the Fresnel integral.

\begin{figure*}
    \centering
    \includegraphics[width=1.\linewidth]{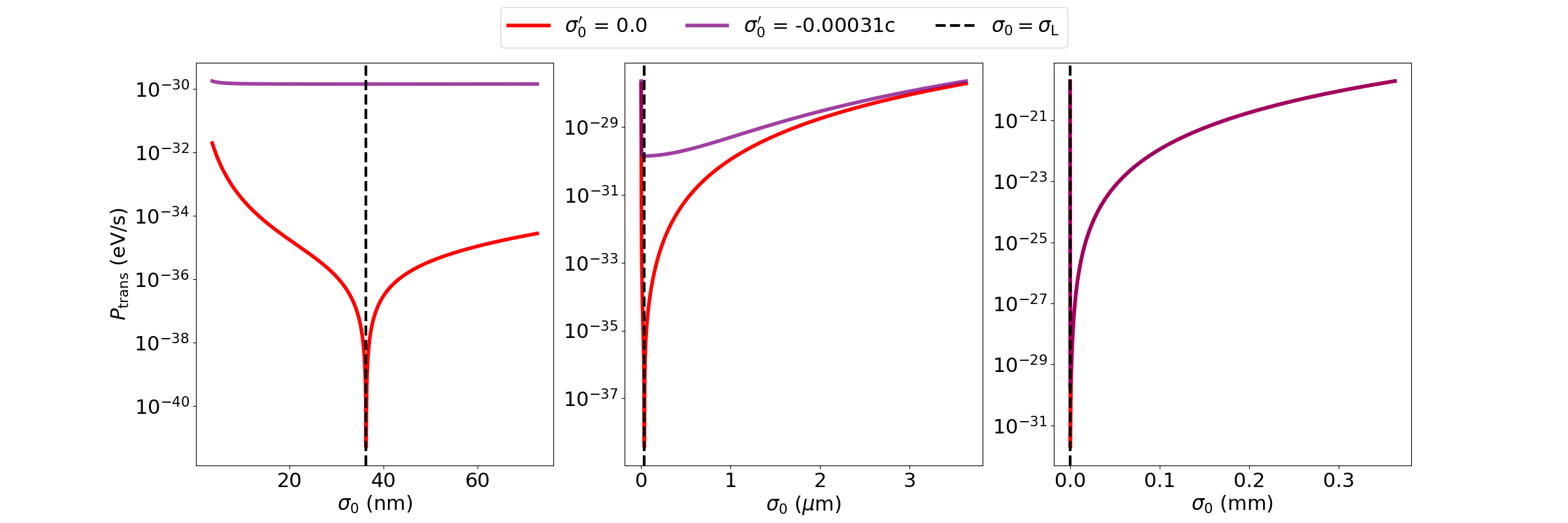}
    \caption{Period-averaged radiated power in the solenoid's fringe region of the characteristic size $D \sim 1 \text{cm}$ as a function of the initial wave packet's width deviation $\sigma_0$. The value of $\sigma_0' = -3.1 \times 10^{-4} c$ was taken from Ref. \cite{schattschneider2014imaging} that has been thoroughly considered in sec. V B of \cite{Sizykh2024Apr}. The following parameters are used: $H = 1$~T (corresponding to $\sigma_{\text{L}} \approx 36$ nm and $\hbar \omega_{\text{c}} \approx 10^{-4}$ eV), $n = 0$, $l = 10$. The panels show the dependence over three distinct scales of $\sigma_0$: (a) nanometers, (b) micrometers, and (c) sub-millimeters.}
    \label{fig:fringe_power}
\end{figure*}

\begin{figure*}
    \centering
    \includegraphics[width=1.\linewidth]{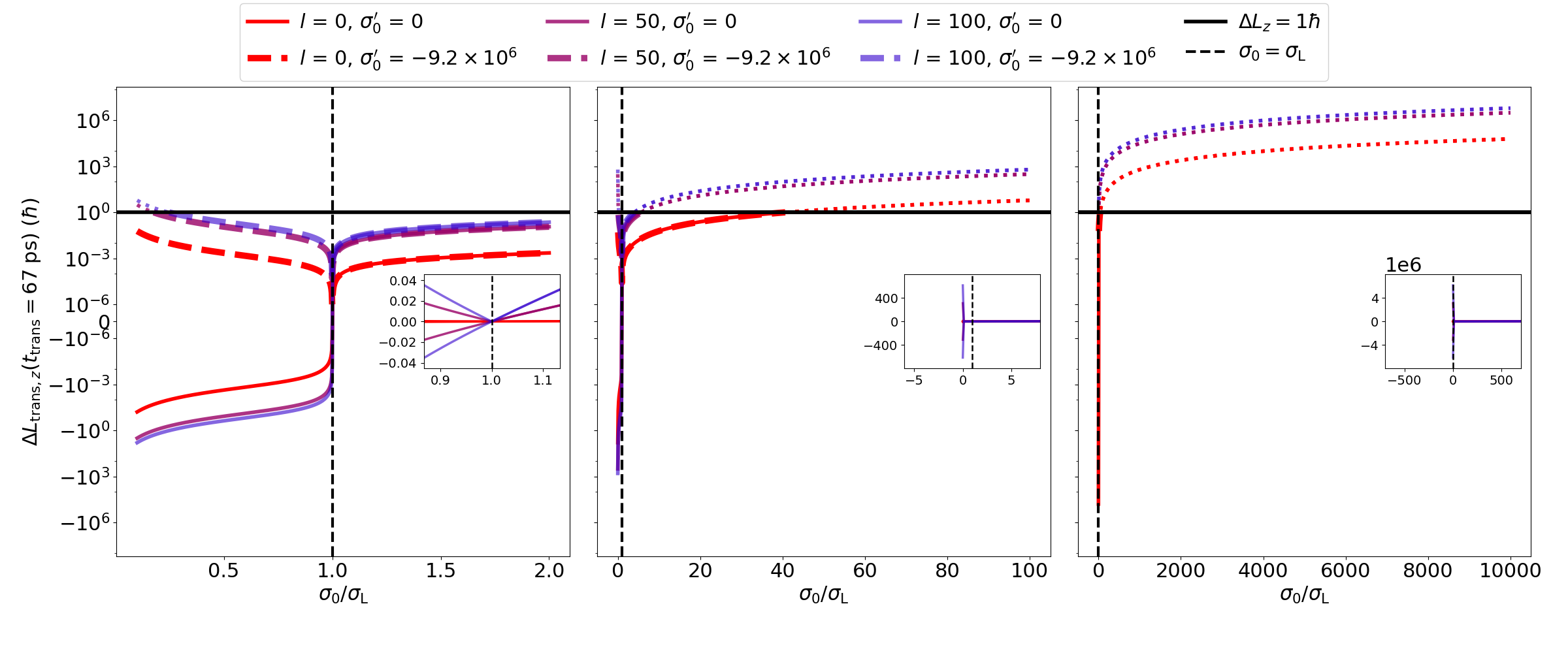}
    \caption{Period-averaged OAM loss per 1 km linac flight time $t = 3.5 \; \mu\text{s}$ in the solenoid's fringe region of the characteristic size $D \sim 1 \text{cm}$ as a function of the normalized initial wave packet's width deviation $\sigma_0/\sigma_\text{L}$. The following parameters are used: $H = 1$~T (corresponding to $\sigma_{\text{L}} \approx 36$ nm and $\hbar \omega_{\text{c}} \approx 10^{-4}$ eV), $n = 0$ and two different values of $\sigma_0'$: $\sigma_0' = 0$ and $\sigma_0' = -3.1 \times 10^{-4} c$ taken from Ref. \cite{schattschneider2014imaging} that has been thoroughly considered in sec. V B of \cite{Sizykh2024Apr}. The panels show the dependence over three distinct scales of $\sigma_0/\sigma_\text{L}$: (a) $\sigma_0 \approx \sigma_\text{L}$, (b) $\sigma_0 > \sigma_\text{L}$, and (c) $\sigma_0 \gg \sigma_\text{L}$. Horizontal line $\Delta L_z = 1 \hbar$ on the graph (c) shows the limitation of the utilized semiclassical approach that does not allow to predict angular momentum loss exceeding the single quantum of OAM. On each figure the subplot of the corresponding scale of $\sigma_0/\sigma_\text{L}$ zoomed to the origin is presented in the linear scale.}
    \label{fig:fringe_oam}
\end{figure*}

We provide the plots of radiation characteristics in figures \ref{fig:fringe_power} and \ref{fig:fringe_oam}, obtained under the assumption that fringe effects occur at the boundary of the solenoid. The number of emitted OAM quanta exceeds 1 for orbital quantum numbers $l \ge 50$ at each considered scale, $\sigma_0/\sigma_\text{L}$, leading to the conclusion that the prepared NSLG state transitions to a state with a different $l$.

Suppose now that the adiabatic approximation does not work. Then, classically, the electron emits the so-called \textit{edge radiation} at a sharp boundary between the drift region and the field of a solenoid. Very similar to transition radiation between two media, there appears a radiation formation length that equals roughly $\gamma^2\lambda$. If the transition region of the solenoid is much smaller than that, then the photon is not formed during that the distance and the fringe field does not make a significant contribution to the radiation power. Thus, for non-relativistic electrons, the fringe fields can indeed be important, but hardly for relativistic ones. If the distance to the next optical element in an accelerator is shorter than the radiation formation length, then the photon is not formed and the edge radiation is suppressed.

Clearly, the fringe fields can result in quantum emission, very similar to Aharonov-Bohm effects, because there is a region where the magnetic field vanishes, wheres the potential does \textit{not}. However, the analysis of such quantum processes lies beyond the scope of our current paper and, moreover, we see no physical reason as to why such quantum emission could significantly increase the emission intensity and the OAM losses.

Note that for the transition region $p_0$ do contribute to the interference term, $\big{\langle} \dd \bm{L}_{p_0} / \dd t \big{\rangle}_{\text{int}, T}$, of OAM change rate. However, in this region the radiative term, $\big{\langle} \dd \bm{L}_{\text{rad}} / \dd t \big{\rangle}_{T}$, remains after averaging over time and provides the leading contribution to the OAM change rate.

\bibliographystyle{apsrev4-2}
\bibliography{references}

\end{document}